\documentclass[12pt]{article}
\begin{document}
\begin{center}
{\bfseries
Hidden Strangeness in Nucleons, Magnetic Moments and SU(3)
}
\vskip 5mm

S.B. Gerasimov$^{ \dag}$
\vskip 5mm
{\small
{\it
Bogoliubov Laboratory of Theoretical Physics\\
Joint Institute for Nuclear Research,
141980 Dubna (Moscow region), Russia}
\\
$\dag$ {\it
E-mail: gerasb@thsun1.jinr.ru
}}
\end{center}
\vskip 5mm
\begin{center}
\begin{minipage}{150mm}
\centerline{\bf Abstract}

The role of nonvalence, e.g., sea quarks and/or
meson) degrees of   freedom in static electroweak
baryon observables is briefly discussed.

\end{minipage}
\end{center}
\vskip 10mm

\section{Introduction}
The constituent quark model is a useful descriptive tool
in hadron spectroscopy.
The empirical spectrum of mesons and baryons suggests that hadrons
are largely composed of the spin-1/2 constituent quarks confined to
 $q \bar q$ and $qqq$ systems.
Naturally, one needs to understand these
relevant degrees of freedom, their effective structure parameters
and forces acting between them to reach the understanding of QCD in the
confining regime.
However, despite many phenomenological successes the fundamental
question of the connection between the three (for baryons)
( or the $q\bar q$-pair, for mesons) "spectroscopic" quarks,
bearing the quantum numbers of a given hadron, and the infinite
number of the "current" (or fundamental) quarks required,
$\it e.g.$, by the deep-inelastic lepton-hadron scattering, is still
a problem.
One can believe, that, among other available menses, the precision
description of static electroweak properties of hadrons, while had been one
of cornerstones of the constituent quark model itself, can still be of
relevance for deeper understanding of fundamental question on the
"constituency" of constituent quarks.
One of aspects of this problem is the following.
The diagonalized Hamiltonian matrix, including the
contributions of all spin-dependent interactions between three
constituent quarks, defines masses of the ground and radially and
orbitally excited states of the nucleon and provides also the structure of
baryon state vectors in the $SU(6)\times O(3)$ basis
states\cite{CI86,Ca92} thus giving the deviation of the $F/D$-values,
parameterizing the electroweak current matrix
elements, from the $F/D=2/3$ that corresponds to participating states in
the $\underline {56}$
-representation, as often assumed for the nucleon and lowest symmetrical
radial excitation (presumably, the Roper resonance with mass $\sim
1440$~Mev).
As discussed in ref.\cite{Ge97}, if the Roper resonance would be
something different from the $3q$-radial excitation, belonging to
the $\underline 56$-representation of $SU(6)$, {\it e.g.} the hybrid
state, then total probability
to find the $(\underline{56})$ - configuration in the nucleon state vector
would be diminished and the $F/D$-value could rise up to the value found
from a version of
phenomenological analysis of the baryon magnetic moments and axial
couplings\cite{Ge95,Ge96}
with important implications for the polarized parton distributions
of the nucleon (see, {\it e.g.},\cite{LSS00} for the recent discussion).

\section{The pionic "dressing" of hadrons: examples and problems}

In this section we consider some consequences from sum rules for the static,
electroweak characteristics of baryons following mainly from the
phenomenology of broken internal symmetries.
We choose the phenomenological sum rule techniques to
obtain, at the price of a minimal number of the  model-dependent
assumptions, a more reliable, though not as much detailed information about
the hadron properties in question.
The main focus will be on the role of nonvalence degrees of
freedom ( the nucleon sea partons and/or peripheral meson currents )
in parameterization and description of hadron magnetic moments, both
diagonal and non-diagonal, including the $N\Delta
\gamma$-transition moment.
Earlier we have considered\cite{Ge95} a number of
consequences of sum rules for the static,
electroweak characteristics of baryons following from the theory of broken
internal symmetries and common features of the quark models
including relativistic effects and corrections due to
nonvalence degrees of freedom -- the sea partons and/or the meson clouds
at the periphery of baryons.\\
Here, we list some of the earlier discussed \cite{Ge95,Ge88,Ge96} sum rules
(we use the particle and quark symbols for corresponding magnetic moments):
\begin{eqnarray}
\alpha_D = {D\over F+D} = {1\over 2}(1-{\Xi^0 - \Xi^- \over \Sigma^+ - \Sigma^-
- \Xi^0 + \Xi^-} \\
{u \over d} = {\Sigma^+ (\Sigma^+ - \Sigma^-) - \Xi^0 (\Xi^0 - \Xi^-) \over
\Sigma^- (\Sigma^+ - \Sigma^-) - \Xi^- (\Xi^0 - \Xi^-)} \\
={P + N + \Sigma^+ - \Sigma^- + \Xi^0 - \Xi^- \over P + N - \Sigma^+ + \Sigma^- - \Xi^0 + \Xi^-}
\end{eqnarray}
The $D$- and $F$- constants in Eq.(1) parameterize "reduced" matrix
elements of quark current operators where $SU(3)$ symmetry-breaking
effects are contained in the factorized effective coupling constants
of the single-quark-type operators, while other contributions
({\it e.g.} representing the pion exchange current effects) are
cancelled in Eqs.(1)-(3) by construction.

The nonzero ratio $u/d$ is related to the chiral constituent quark
model where a given baryon consists of three "dressed", massive
constituent quarks. Owing to the virtual transitions
$q \leftrightarrow  q + \pi(\eta), q \leftrightarrow K + s$
the "magnetic anomaly" is developing, {\it i.e.,}
$u/d = -1.80 \pm .02 \neq Q_{u}/Q_{d}=-2$. Evaluation of the one-loop
quark-meson diagrams done earlier \cite{Ge88}
gives: $u/d = (Q_{u}+\kappa_{u})/(Q_{d}+\kappa_{d}) \simeq -1.85$,
the $\kappa_{q}$ being the quark anomalous magnetic moment in natural units,
if we take the $SU(3)$-invariant quark-pseudoscalar-meson couplings,
the physical masses for the $\pi$-, $\eta$-, K-mesons and the
$m_{q(s)} \simeq 300(400)$ MeV. In the full $SU(3)$-symmetry limit, when
$m_q = m_s, m_{\pi} = m_{\eta} = m_K$, we return to $u/d = -2$, the ratio
pertinent to the structureless current quarks. Evaluation of the "magnetic
anomaly" via Eq.(3) turns out rather close to the result of the lowest order
quark-pion loop diagrams with the pseudoscalar $qq\pi$-coupling estimated
via the Goldberger-Treiman relation at the quark level \cite{Ge79}, while
the terms of order $O(g^{4}_{qq\pi})$ are negligible if the use is made
of old calculations for magnetic moments of nucleons \cite{Na54}
to adapt them for the quark case through corresponding replacement of
the coupling constants and masses of fermions.
The ratio $s/d \simeq .64$ demonstrating the $SU(3)$-symmetry breaking
is evaluated via
\begin{eqnarray}
{s \over d} = {\Sigma^+ \Xi^- - \Sigma^- \Xi^0 \over \Sigma^- (\Sigma^+ - \Sigma^-)
- \Xi^- (\Xi^0 - \Xi^-)}
\end{eqnarray}
Now, we list some consequences of the obtained sum rules.
The numerical relevance of adopted parameterization is seen from results
enabling even  to estimate from one of obtained sum rules, namely,
\begin{eqnarray}
&&(\Sigma^+ - \Sigma^-) (\Sigma^+ + \Sigma^- - 6\Lambda + 2\Xi^0 + 2\Xi^-) \nonumber \\
&&- (\Xi^0 - \Xi^-) (\Sigma^+ + \Sigma^- + 6\Lambda - 4\Xi^0 - 4\Xi^-) = 0.
\end{eqnarray}
the necessary effect of the isospin-violating $\Sigma^{o} \Lambda$-mixing.
By definition, the $\Lambda$--value
entering into Eq.(18)
should be "refined" from the
electromagnetic $\Lambda\Sigma^0$--mixing affecting  $\mu(\Lambda)_{exp}$.
Hence, the numerical value of $\Lambda$, extracted from Eq.(18),
can be used to determine the $\Lambda\Sigma^0$--mixing angle through the
relation
\begin{eqnarray}
\sin \theta_{\Lambda\Sigma} \simeq \theta_{\Lambda\Sigma} =
{\Lambda -\Lambda_{exp} \over 2\mu(\Lambda\Sigma)} = (1.43 \pm 0.31) 10^{-2}
\end{eqnarray}
in accord with the independent estimate of $\theta_{\Lambda\Sigma}$ from
the electromagnetic mass-splitting sum rule \cite{DvH}.
Of course, this approach is free of a problem raised by Lipkin\cite{Li99}
and concerning the ratio $R_{\Sigma / \Lambda}$
of magnetic moments of $\Sigma$- and $\Lambda$- hyperons.
With the parameters
$u/d=-1.80$ and $\alpha_{D}=(D/(F+D))_{mag}=0.58$, defined without
including in fit the $\Lambda$-hyperon magnetic moment, we obtain
\begin{equation}
R_{\Sigma / \Lambda} = \frac{\Sigma^{+}+2\Sigma^{-}}{\Lambda}
= -.27~~ (\mbox{\it vs}~~ -.23~ \cite{PDG})
\end{equation}
while in the standard nonrelativistic quark model
without inclusion of nonvalence d.o.f. this ratio would equals $-1$.

The experimentally interesting quantities $\mu(\Delta^+ P) = \mu(\Delta^0 N)$
and $\mu(\Sigma^{*0}\Lambda)$ are affected by the exchange current
contributions and for their estimation we need additional assumptions.
We use the analogy with the one--pion--exchange current, well--known in
nuclear physics, to assume for the exchange magnetic moment operator
\begin{eqnarray}
\hat\mu_{exch} =\sum\limits_{i<j} [\vec\sigma_i \times \vec\sigma_j]_3
[\vec\tau_i \times \vec\tau_j]_3 f(r_{ij}),
\end{eqnarray}

where $f(r_{ij})$ is a unspecified function of the interquark distances,
$\vec\sigma_i (\vec\tau_i)$ are spin (isospin) operators of quarks.
Calculating the matrix elements of $\hat\mu_{exch}$ between the baryon wave
functions, belonging to the 56--plet of $SU(6)$, one can find

\begin{eqnarray}
&& \mu(\Delta^+P)_{\underline 56} = {1\over \sqrt{2}} \left(P - N +{1\over
3} (P+N) {1-u/d\over 1+ u/d}\right).
\end{eqnarray}
where Eq.(24) may serve as a generalization of the well--known
$SU(6)$--relation \cite{GR64,BLP64}.

We list below the limiting relations following
from the neglect of the nonvalence degrees of freedom

\begin{eqnarray}
\Sigma^+ [\Sigma^-] = P[-P-N] +
(\Lambda - {N\over 2})( 1+{2N \over P}),\\
\Xi^0[\Xi^-] = N[-P-N] + 2(\Lambda -
{N\over 2})( 1+ {N\over 2P}),\\
\mu(\Lambda\Sigma) = -{\sqrt{3}\over 2}N.
\end{eqnarray}

The numerical values of magnetic moments following from this assumption
coincide almost identically with the
results of the $SU(6)$--based NRQM taking account of the $SU(3)$ breaking
due to the quark--mass differences \cite{Ge65}. We stress, however, that no
NR assumption or explicit $SU(6)$-wave function are used this time.
The ratio $\alpha_{D}=D/{(F+D} = .61$ in this case and it is definitely
less than
$ \alpha_{D}=.58$,~ when nonvalence degrees of freedom are
included\cite{Ge95}.
This is demonstrating a substantial influence of the nonvalence
degrees of freedom on this important parameter.

\section
{The OZI-Rule and SU(3) Symmetry
Violation in Magnetic and Axial Couplings of Baryons}

Here, we follow a complementary view of the nucleon structure,
keeping the constraint u/d=-2,and the OZI-rule violating contribution
of sea quarks, parameterized as \\
$\Delta (N)= \sum_{q=u,d,s} \mu(q)<N|\bar{s}s|N> \neq 0 $.

We shall refer to this approach \cite{Ge95} as a correlated current-quark picture
of nucleons. We have then the following important sum rules
(in n.m.)

\begin{eqnarray}
 \Delta(N) = {1\over 6}(3(P+N) - \Sigma^+  + \Sigma^- -\Xi^0 +\Xi^-)&=&
-.06\pm.01,\\
 \mu_N(\overline ss)=\mu(s)
\langle N|\overline ss|N\rangle=
(1- {d\over s})^{-1}\Delta(N) &=& .11,
\end{eqnarray}

where the ratio d/s=1.55 follows
from the correspondingly modified Eq.(4) (that is with $Y$ replaced by $Y - \Delta)$).
By definition, $\mu_N(\overline ss)$ represents the contribution of
strange ("current") quarks to nucleon magnetic moments.
Numerically, our $\mu_N(\bar{s}s$
agrees fairly well with other more specific models ( see,e.g. \cite{Mu94}.
Actually, our Eqs. (13) and (14) are equivalent, up to the common factor
$-(1/3)$, which is the electric charge of the strange quark, to the half-sum
of two relations in Ref.\cite{Le98} that refer to $\mu_N(\bar{s}s)$ and
where the ratios of effective magnetic moments of quarks in different baryons
should be taken the same. We stress that the sign of our $\mu(\bar{s}s)$
is opposite to one of the central measured value in the experiment
SAMPLE \cite{Mue97}.

The calculated quantity indicates violation of the OZI rule and the strange
current quark contribution $\mu_N(\overline ss)$ is seen to constitute
a sizable part of the isoscalar magnetic moment of nucleons (or, which is
approximately the same, of the nonstrange constituent quarks)
\begin{eqnarray}
{1\over 2}(P+N)=\mu(\overline uu + \overline dd)
+\mu(\overline ss) =.44,
\end{eqnarray}
This observation helps to understand the unexpectedly
large ratio
\begin{eqnarray}
BR\left ( {{\overline PN\to \phi +\pi}
\over {\overline PN\to \omega +\pi}}\right )\simeq(10\pm2)\% ,
\end{eqnarray}
reported for the s--wave $\bar NN$ -- annihilation reaction \cite{Fa93}.

Indeed, the transition ${(\overline PN)}_{s-wave} \to V+\pi$,
where $V=\gamma,\omega,\phi$, is of the magnetic dipole type. Therefore,
the transition operator should be proportional to the isoscalar
magnetic moment contributions
from the light u-- and d--quarks and
the strange s--quark, Eq.(25). The transition
operators for the $\omega$-- and $\phi$--mesons are obtained from
$\mu(\overline qq)$ and $\mu(\overline ss)$  through the well--known
vector meson dominance model ( VDM ).Using  the "ideal" mixing ratio
$g_\omega:g_\phi=1:\sqrt 2$ for the
photon--vector--meson junction couplings
and $\mu_\omega:\mu_\phi=\mu(\overline qq):\mu(\overline ss)$
we get
\begin{eqnarray}
BR\left ({{ \overline PN \to \phi +\pi} \over
{\overline PN \to \omega +\pi}}\right )
\simeq \left ({ \mu( \overline ss)} \over
{\sqrt 2 \mu( \overline uu + \overline dd)}
\right )^2 \left ({p_\phi ^{c.m.}}
\over {p_\omega^{c.m.}} \right )^3 \simeq  6\%,
\end{eqnarray}
which is reasonably compared with data.

To estimate possible influence of the SU(3) breaking in the ratio of the
weak axial-to-vector coupling constants we adopt the following
prescription suggested by the success of our parameterization of the
baryon magnetic moment values within constituent quark model. In essence,
we assume that the leading symmetry breaking effect is
produced by different renormalization
of the $\bar q q W$- strangeness-conserving and strangeness-nonconserving
vertices with the participation of the constituent quarks.
As to the baryon wave functions and the transition matrix
elements of the standard octet (i.e. Cabibbo) currents, they will
fulfill separately , within the strangeness-conserving or, respectively,
the stranegeness-nonconserving coupling constants. Hence we have two sets
of the $SU(3)$-symmetry relations but, in general, with different values of
$F$- and $D$-type coupling constants in each set.

So that, to obtain the contributions of the u-,d-,
and s-flavoured quarks to the "proton spin" (or, rather to certain
combination of the axial current matrix elements),
denoted by $\Delta u(p), \Delta d(p)$ and $\Delta s(p)$, the use
should be made of baryon semileptonic weak decays with $(\Delta S=0)$,
treated with the help of the exact $SU(3)$-symmetry. It has been shown
in Ref.\cite{Ra96}
that when both the strangeness-changing
$(\Delta S=1)$ and strangeness-conserving
$(\Delta S=0)$ transitions are taken for the analysis,
then $(D/F+D)_{ax}^{\Delta S=0,1} = .635 \pm .005 $ while
$(D/F+D)_{ax}^{\Delta S=0}=.584 \pm .035 $ (which is more close to
$ (D/D+F)_{mag}$).

We list below two sets of the $\Delta q$-values,
we have obtained from the data
with inclusion of the QCD radiative corrections
(e.g.\cite{Ad94} and references
therein) : $\Delta u(p) \simeq .82 (.83),\;
\Delta d(p) \simeq -.44(-.37),\;
\Delta s =-.10 \pm .04(-.19 \pm .05)$, where the values in
the parentheses correspond to $\alpha _{D}=(D /D+F)=.58$.
Concerning the very $g_a(\Delta S=1)$-constants, we mention
the following options. In all but one \cite{Hs88} analyses of
the hyperon $\beta$-decays,  the absence of the "weak electricity"
form factor $g_2(Q^2)$ due to induced second class weak current
has been postulated from the very beginning.
Following this way and taking $g_{a}^{exp}(\Sigma^{-} \rightarrow
n)=-.34 \pm .024$, one gets \cite{Ra96} $(F/D)_{\Delta S=1}=.575 $,
demonstrating difference from $(F/D)_{\Delta S=0}=.72$ and the
different tendency of deviation from $(F/D)_{SU(6)}=2/3$, the fact
which has to be understood dynamically. The fit to all
$\Sigma^{-} \rightarrow ne\bar {\nu}$ decay data of Ref. \cite{Hs88}
 with $g_2 \neq 0$ yields $g_a=-.20 \pm .08$ and $g_2=+.56 \pm .37$.
 One cannot then define $(F/D)_{\Delta S=1}$ because data for all other
 hyperons have been treated under the assumption $g_2=0$.
 However, noting close numerical values of two quantities
\begin{eqnarray}
 F - D = g_a^{exp}(\Sigma^{-}\rightarrow n) = -.20 \pm .08, \nonumber \\
 F - D = g_a^{exp}(n \rightarrow p) - \sqrt6
g_a^{exp}(\Sigma \rightarrow \Lambda)= -.19 \pm .04,\nonumber
\end{eqnarray}
one can suggest, in this particular case, the universal value
$(F/D)_{\Delta S=0,1}=.72$ to get
\begin{eqnarray}
g_a(\Lambda \rightarrow p) = F + \frac{1}{3}D = .77;~~ (.718 \pm .015),\\
g_a(\Xi^{-} \rightarrow \Lambda) = F - \frac{1}{3} D = .28;~~ (.25 \pm .05),\\
g_a(\Xi^{o} \rightarrow \Sigma^{+}) = F + D = 1.26;~~ (1.32 \pm .20)
\end{eqnarray}
where the values in parentheses have been derived \cite{PDG,Al01} from the
analyses of data with the additional constraint $g_2=0$. The most significant
difference in two sets of $g_a$ values is seen for the $\Lambda \rightarrow
p e \bar {\nu}$ decay which, therefore, lends itself as the best candidate
for an alternative treatment of available or improved new data with
the "weak-electricity" $g_2$- constant taken as a free parameter to be
determined from data simultaneously with the axial-vector and weak-magnetism
constants.

\section{Concluding remarks}

Besides importance of resolution of the problem on the presence
and quantitative role of the weak second-kind current and
corresponding form factors in the hyperon $\beta$-decay
observables, one can mention also about major theoretical
interest in the careful study of the strangeness-conserving
$\Sigma ^{\pm} \rightarrow \Lambda e^{\pm} \nu(\bar{\nu})$
transitions which would not only prove (or disprove) hypotheses
about dependence of $(F/D)$-ratios on the $\Delta S$, labelling
the transitions, but also would provide information on the isospin
breaking effects underlying the $\Lambda - \Sigma^{o}$ - mixing.

There is presently well-justified experimental and theoretical interest
concerning the question of hidden strangeness in the nucleon.
The SAMPLE collaboration has recently reported the
first measurement of the strange magnetic moment of the
proton~\cite{Mue97}, which turned out very close to zero, but with
rather large experimental errors, "touching" both significant positive
values of $\mu_{s}$ (in our normalization), as majority models
also predict,
as well as the opposite sign value, claimed mainly within chiral
soliton models ({\it {e.g.}} \cite {Ho01, Pr01},
and references therein).
For a detailed understanding about
the strength of the various strange operators in the proton
one has to wait until the dedicated programs at JLab,
BATES (MIT) and MAMI (Mainz) will be fulfilled
on the measurement of parity violation in the helicity dependence of
electron-nucleon scattering in wider range of the momentum-transfers to
separate the contributions of the charge, magnetic and axial terms,
thus providing data about the basic quark structure of the nucleon.

\section*{Acknowledgments}
The author would like to express his deep gratitude to
Prof. T. Morii for organization of
the Workshop  and for most warm hospitality extended to him.
This work was also supported in part by the Russian Foundation for Basic
Research grant No. 00-15-96737.

\end{document}